\documentclass[aps,pra,twocolumn,showpacs,showkeys,groupedaddress,nobalancelastpage,nofootinbib,floatfix,superscriptaddress,longbibliography]{revtex4-2}
\usepackage{graphics}
\usepackage{graphicx}% Include figure files
\usepackage{dcolumn}% Align table columns on decimal point
\usepackage{bm}% special 'bold-math' package
\usepackage{comment} %comment multiple lines
\usepackage{xcolor}
\usepackage{bookmark}
\usepackage{tensor}
\usepackage{amsmath}
\usepackage{amssymb}
\usepackage{float}
\usepackage[utf8]{inputenc}
\usepackage{romannum}
\usepackage{hyperref}
\usepackage[capitalize]{cleveref}
\crefname{section}{Sec.}{Sec.}

\begin{document}
\pagenumbering{arabic}
%%%%%%%%%%%%%%%%%%%%%%%
% Additional Commands %
%%%%%%%%%%%%%%%%%%%%%%%
\newcommand\be{\begin{equation}}
\newcommand\ee{\end{equation}}
\newcommand\bea{\begin{eqnarray}}
\newcommand\eea{\end{eqnarray}}
\newcommand\ket[1]{|#1\rangle}
\newcommand\bra[1]{\langle #1|}
\newcommand\braket[2]{\langle #1|#2\rangle}
%%%%%%%%%%%%%%%%%%%%%%%%%%%%%%%%%%
\def\dbar{{\mathchar'26\mkern-12mu d}} %inexact differential symbol
%%%%%%%%%%%%%%%%%%%%%%%
% Theorem Environment %
%%%%%%%%%%%%%%%%%%%%%%%
 \newtheorem{thm}{Theorem}
 \newtheorem{cor}[thm]{Corollary}
 \newtheorem{lem}[thm]{Lemma}
 \newtheorem{prop}[thm]{Proposition}
 \newtheorem{defn}[thm]{Definition}
 \newtheorem{rem}[thm]{Remark}
%%%%%%%%%%%%%%%%%%%%%%%%%%%%%

%%%%%%%%%%%%%%%%%%%%%
% Title and Authors %
%%%%%%%%%%%%%%%%%%%%%
\pagestyle{plain}
\title{\bf A Quantum Otto Engine with Shortcuts to Thermalization and Adiabaticity}
\author{ A. Pedram}
\email{apedram19@ku.edu.tr}
\affiliation{Department of Physics, Koç University, Istanbul, Sarıyer 34450, Türkiye}
\author{ S. C. Kadıoğlu}
\email{skadioglu16@ku.edu.tr}
\affiliation{Department of Physics, Koç University, Istanbul, Sarıyer 34450, Türkiye}
\author{A. Kabakçıoğlu}
\email{akabakcioglu@ku.edu.tr}
\affiliation{Department of Physics, Koç University, Istanbul, Sarıyer 34450, Türkiye}
\author{Ö. E. Müstecaplıoğlu}
\email{omustecap@ku.edu.tr}
\affiliation{Department of Physics, Koç University, Istanbul, Sarıyer 34450, Türkiye}
\affiliation{TÜBİTAK Research Institute for Fundamental Sciences, 41470 Gebze, Türkiye}
\pagebreak

%%%%%%%%%%%%
% Abstract %
%%%%%%%%%%%%
\begin{abstract}
  We investigate the energetic advantage of accelerating a quantum harmonic oscillator (QHO) Otto engine by use of shortcuts to adiabaticity (for the expansion and compression strokes) and to equilibrium (for the hot isochore), by means of counter-diabatic (CD) driving. By comparing various protocols with and without CD driving, we find that, applying both type of shortcuts leads to enhanced power and efficiency even after the driving costs are taken into account. The hybrid protocol not only retains its advantage in the limit cycle, but also recovers engine functionality (i.e., a positive power output) in parameter regimes where an uncontrolled, finite-time Otto cycle fails. We show that controlling three strokes of the cycle leads to an overall improvement of the performance metrics compared with controlling only the two adiabatic strokes. Moreover, we numerically calculate the limit cycle behavior of the engine and show that the engines with accelerated isochoric and adiabatic strokes display a superior power output in this mode of operation. 
\end{abstract}
\maketitle

%%%%%%%%%%%%%%%%%%%
% The Article Body%
%%%%%%%%%%%%%%%%%%%
\section{Introduction}
\label{sec:intro}
Enhancing the performance of heat engines has been a long-standing objective in the field of thermodynamic cycle investigations. The pinnacle of these endeavors is reflected in Carnot's second law of thermodynamics, which establishes an upper limit on the efficiency of heat engines~\cite{bhlitem57177}. This limit can, in principle, be attained through a Carnot cycle, comprising reversible processes. However, while the Carnot engine achieves this limit, it is unable to generate any power due to the quasi-static nature of its processes. Practical heat engine cycles such as Otto, Diesel, or Stirling face the challenge of balancing efficiency and power, necessitating real-world engines to complete a cycle within a finite time, albeit with lower efficiency. Of particular significance is the efficiency at maximum power, which was studied by Curzon and Ahlborn in their seminal paper~\cite{doi:10.1119/1.10023}.\\
Currently, there is ongoing research in quantum heat engines and refrigerators~\cite{doi:10.1146/annurev-physchem-040513-103724,Kosloff2017TheQH,tuncer2020,kurizki_kofman_2022,10.1088/2053-2571/ab21c6}. Since the pioneering work by Scovil and Schulz-DuBois~\cite{PhysRevLett.2.262}, quantum heat engines have been extensively studied theoretically~\cite{1979,doi:10.1063/1.446862,doi:10.1063/1.461951,10.1063/1.471453,10.1119/1.18197,PhysRevLett.88.050602,PhysRevLett.89.116801,doi:10.1126/science.1078955,PhysRevE.68.016101,PhysRevLett.93.140403,PhysRevE.70.046110,PhysRevE.72.056110,Rezek_2006,PhysRevE.76.031105,PhysRevE.77.041118,PhysRevLett.109.203006,PhysRevE.90.022102,Hardal2015,PhysRevE.95.032139,Turkpence_2017,PhysRevB.96.104304,PhysRevE.96.062120,Niedenzu2018,PhysRevE.97.062153,Naseem:19,PhysRevE.100.012109,PhysRevE.102.062123,Hamedani_Raja_2021,e24101458,Pozas-Kerstjens_2018,PhysRevA.106.L030201}, and successful experimental demonstrations have been recently reported~\cite{PhysRevLett.119.050602,PhysRevLett.122.110601,PhysRevLett.123.240601,Bouton2021,doi:10.1126/sciadv.abl7740,PhysRevA.106.022436,Zhang2022,e25020311}. Despite these theoretical and experimental developments, it is still an open question whether the quantum advantage can lead to Carnot efficiency or better at finite power~\cite{PhysRevLett.112.030602,Campisi2016,Kim2022}. It is, therefore, crucial to study the potential and limitations of quantum thermal devices and investigate methods to boost their finite-time performance.\\

A specific type of control techniques which are collectively known as shortcuts to adiabaticity (STA) have gained a particular interest in the study of quantum heat engines. The central idea in STA is to design protocols to drive the system by emulating its adiabatic dynamics in finite time~\cite{TORRONTEGUI2013117,RevModPhys.91.045001}. Since its foundation~\cite{doi:10.1021/jp030708a,doi:10.1021/jp040647w,Berry_2009,PhysRevLett.104.063002} different approaches have been developed to engineer STA, such as using dynamical invariants,~\cite{PhysRevLett.104.063002,PhysRevA.83.062116}, inversion of scaling laws~\cite{PhysRevA.84.031606,Campo2012} and the fast-forward technique~\cite{doi:10.1098/rspa.2009.0446,PhysRevA.84.043434,PhysRevA.86.013601}. In recent years, successful experimental realizations of STA have been reported in the literature~\cite{doi:10.1126/sciadv.aau5999,https://doi.org/10.1002/qute.201900121,PhysRevApplied.13.044059,Yin2022}.\\

STA has been extensively explored as a means to enhance the performance of quantum refrigerators~\cite{PhysRevB.100.035407,PhysRevResearch.2.023120} and boost the power output of QHO~\cite{Campo2014,PhysRevE.98.032121,PhysRevE.99.022110,e18050168} and spin chain~\cite{PhysRevE.99.032108,PhysRevResearch.2.023145,_AKMAK_2021} based quantum heat engines by emulating adiabatic strokes within finite time frames. Previous studies in these domains have commonly assumed a rapid thermalization step. However, with the growing interest in the application of STA to open quantum systems, recent research has increasingly focused on the possibility of achieving swift thermalization using various techniques~\cite{PhysRevA.98.052129,PhysRevLett.122.250402,PhysRevA.100.012126,Alipour2020shortcutsto,PhysRevResearch.2.033178,PhysRevX.10.031015,Suri2018,Dann_2020}. These approaches are often referred to as fast thermalization, shortcuts to thermalization (STT) or shortcuts to equilibrium (STE). Since quantum heat engines contain thermalization branches, study of these techniques is of paramount importance in order to optimize the power and efficiency of these devices~\cite{PhysRevA.100.012126,PhysRevX.10.031015,Suri2018,Dann_2020,delCampo2018}.\\

Motivated by these advancements, we conduct an analysis of the energy dynamics, power output, and efficiency characteristics of an Otto cycle utilizing a QHO as its working medium within a finite time frame. To achieve this, we employ shortcuts to adiabaticity (STA) during the expansion and compression strokes, while implementing STE during the hot isochoric strokes, resulting in what we refer to as an ST{\AE} engine. In our investigation, we compare the thermodynamic performance metrics of this engine with an uncontrolled non-adiabatic quantum Otto (UNA engine) and a quantum Otto engine in which only the adiabatic strokes are accelerated using CD-driving (STA engine).\\

For a comprehensive evaluation of the thermodynamic performance within controlled dynamics, it is essential to meticulously account for the energetic expenses associated with the implementation of the control protocols~\cite{PhysRevE.99.032108,PhysRevE.98.032121,PhysRevE.99.022110,PhysRevResearch.2.023145}. In this regard, we adopt a widely employed cost measure to quantify the energetic requirements of the STA protocol during the adiabatic strokes. Additionally, we introduce a novel cost measure based on the dissipative work performed by the environment~\cite{PhysRevA.105.L040201} specifically for STE protocol. This approach allows for an accurate evaluation of energy-related aspects when both STA and STE protocols are implemented.\\

Initially, we conduct a thermodynamic analysis of the three engines (UNA, STA, and ST{\AE}) by considering a single cycle following the preparation phase. Subsequently, we extend our examination to their respective limit cycles. Our findings clearly demonstrate that the ST{\AE} engine, incorporating both shortcuts to adiabaticity (STA) and shortcuts to equilibrium (STE) protocols, attains superior power output and efficiency when compared to the STA-only engine and the UNA engine.\\

This manuscript is organised as follows. In~\cref{sec:otto} we briefly introduce the quantum Otto cycle. In~\cref{sec:sta} the scheme for STA using counterdiabatic driving is introduced. In~\cref{sec:stt} the protocol for the fast driving of the Otto cycle towards thermalization is introduced. In~\cref{sec:otto_sta_ste} we describe the quantum Otto engine for which both STA and STE is used. In~\cref{sec:result} we present our results and finally in~\cref{sec:conclusion} we draw our conclusions.

\section{Quantum Harmonic Otto Cycle}
\label{sec:otto}
\begin{figure}[htbp!]
\centering
\includegraphics[width=\linewidth]{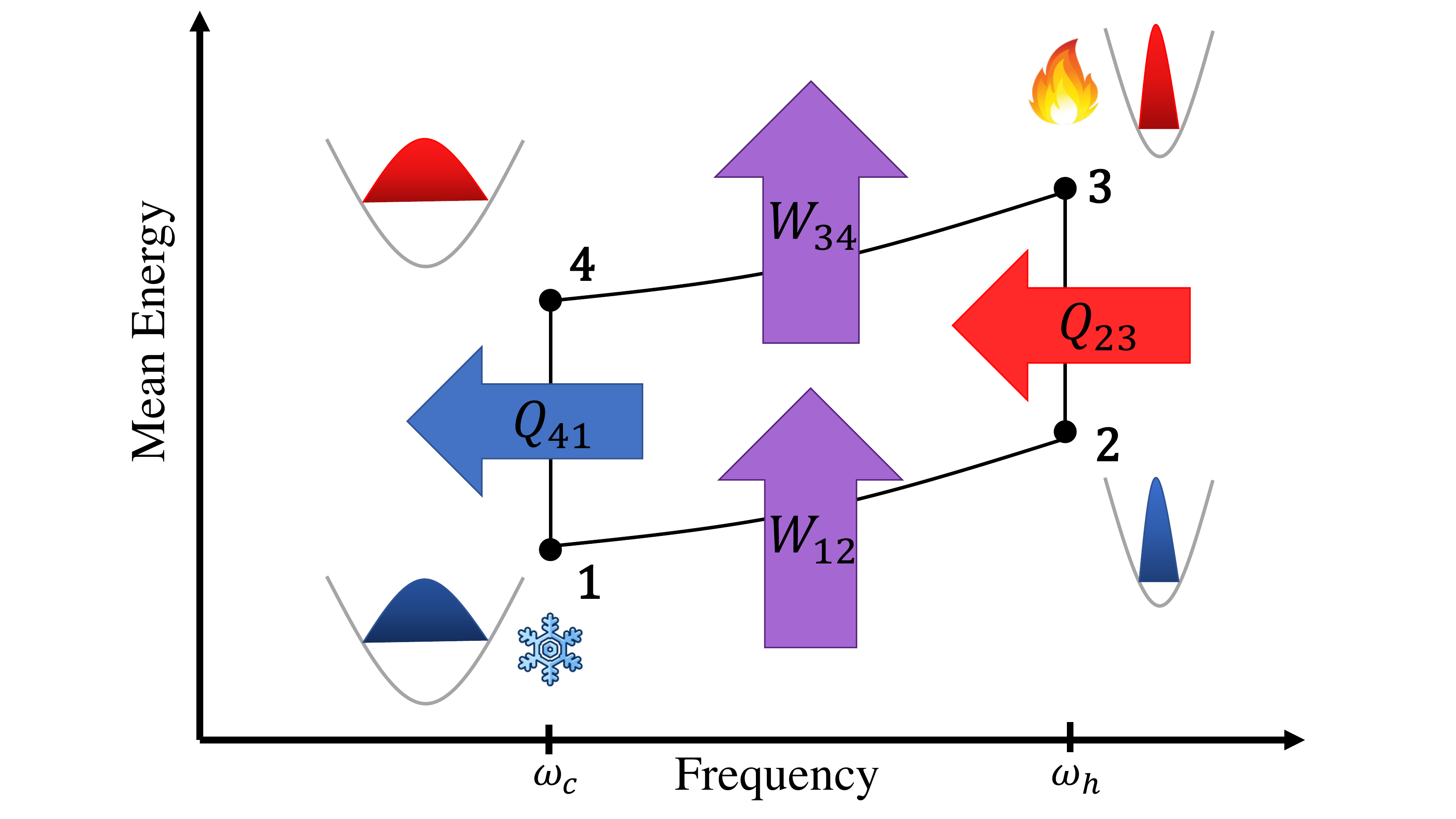}
\caption{Schematic representation of a quantum Otto cycle composed of adiabatic compression ($1\rightarrow2$), isochoric hot thermalization ($2\rightarrow3$), adiabatic expansion ($3\rightarrow4$) and isochoric cold thermalization ($4\rightarrow1$) strokes. \label{otto.schm}}
\end{figure}

We study a finite-time quantum Otto cycle and consider the working medium to be a harmonic oscillator with a time dependent frequency. The Hamiltonian for a QHO is 
\begin{equation}\label{qhohamiltonian}
  \hat{H}_0=\hat{p}^2/2m +m\omega_t^2\hat{x}^2/2
\end{equation}
in which $\hat{x}$ and $\hat{p}$ are the position and momentum operators respectively and $\omega_t$ is the time-dependent frequency and $m$ is the mass of the oscillator. As shown in \cref{otto.schm}, the cycle consists of the following strokes.
\begin{itemize}
\item[(1)] Adiabatic compression ($1\rightarrow 2$): Unitary evolution
  for a duration of $\tau_{12}$ while the frequency increases
  from $\omega_c$ to $\omega_h$.
\item[(2)] Hot isochore ($2\rightarrow 3$): Heat transfer from the hot
  bath to the system for a duration of $\tau_{23}$ as the system
  equilibrates with the bath at temperature $T_h$ at fixed frequency.
\item[(3)] Adiabatic expansion ($3\rightarrow 4$): Unitary evolution
  for a duration of $\tau_{34}$ while the frequency decreases
  from $\omega_h$ to $\omega_c$.
\item[(4)] Cold isochore ($4\rightarrow 1$): Heat transfer from the
  system to the cold bath for a duration of $\tau_{41}$ as the system
  equilibrates with the bath at temperature $T_c$ at fixed frequency.
\end{itemize}

During the adiabatic steps the system is decoupled from the heat baths, and the frequency is varied over time, at a rate bounded by the adiabatic theorem~\cite{Berry_2009}, so that the populations of the drifting energy levels remain unchanged. The resulting isentropic evolution is more restrictive than the classical adiabaticity condition of constant entropy. This is due to the fact that in a quantum adiabatic evolution, the adiabatic condition implies that the distribution of the populations must remain unchanged. However, in a classical adiabatic process the entropy is unchanged~\cite{PhysRevE.76.031105}.

Mathematical formulations of the adiabatic time evolution for an isolated quantum system employ a time-scale separation between fast and slow degrees of freedom, yielding product states as solutions. For the compression and expansion strokes of the QHO Otto engine, this condition implies that the system will always remain in an instantaneous eigenstate of the time-dependent Hamiltonian. The unitary time evolution of such a process, which ideally takes infinite time, can be solved exactly and has zero entropy production. The average work and heat exchange values during the cycle is given by~\cite{PhysRevE.77.021128,DEFFNER2010200},
\begin{eqnarray}\label{adiabatic.cycle}
   \langle W_{12}\rangle &=& \frac{\hbar}{2}(\omega_h Q_{12}^{*} - \omega_c) \text{coth}(\frac{\beta_c \hbar \omega_c}{2}) \\
   \langle Q_{23}\rangle &=& \frac{\hbar \omega_h}{2} \left[\text{coth}(\frac{\beta_h \hbar \omega_h}{2}) - Q_{12}^{*}\text{coth}(\frac{\beta_c \hbar \omega_c}{2}) \right] \\
   \langle W_{34}\rangle &=& \frac{\hbar}{2}(\omega_c Q_{34}^{*} - \omega_h) \text{coth}(\frac{\beta_h \hbar \omega_h}{2}) \\
   \langle Q_{41}\rangle &=& \frac{\hbar \omega_c}{2} \left[\text{coth}(\frac{\beta_c \hbar \omega_c}{2}) - Q_{34}^{*}\text{coth}(\frac{\beta_h \hbar \omega_h}{2}) \right]
\end{eqnarray}
where $W$ refers to the work input and $Q$ refers to the heat input to the engine during the indicated strokes. $\beta_c$ and $\beta_h$ are the inverse temperatures for the cold and hot bath respectively. The terms $Q^*_{12}$ and $Q^*_{34}$ are the adiabaticity parameters and their values depend on the driving scheme~\cite{PhysRevE.98.032121}. We are employing a sign convention in which all the incoming fluxes (heat and work) are taken to be positive~\cite{PhysRevE.106.024137,PhysRevE.102.062123}.\\

In order to study the finite-time Otto cycle, we need to consider the time evolution along each stroke separately. The unitary evolution of the adiabatic strokes is governed by
\begin{equation}\label{unitary.dynamics}
   \partial_t\hat{\rho} = -\frac{i}{\hbar}[\hat{H}_0,\hat{\rho}],
\end{equation}
where $\hat{\rho}$ is the system's density matrix. For the duration of the isochoric strokes, a Markovian open system dynamics given by a Gorini–Kossakowski–Sudarshan–Lindblad (GKLS) master equation governs the system's dynamics,
\begin{equation}\label{lindblad.dynamics}
   \partial_t\hat{\rho} = -\frac{i}{\hbar}[\hat{H}_0,\hat{\rho}] + \mathcal{D}(\hat{\rho}),
\end{equation}
in which $\mathcal{D}(\hat{\rho})$ is the dissipation superoperator which must conform to Lindblad’s form for a Markovian evolution
 \begin{equation}\label{lindblad.diss}
   \mathcal{D}(\hat{\rho}) =  k_{\uparrow } (\hat{a}^\dagger \hat{\rho} \hat{a} - \frac{1}{2}\{\hat{a} \hat{a}^\dagger, \hat{\rho}\})+ k_{\downarrow } (\hat{a} \hat{\rho} \hat{a}^\dagger - \frac{1}{2}\{\hat{a}^\dagger \hat{a}, \hat{\rho}\}).
 \end{equation}
The $ k_{\uparrow }$ and $ k_{\downarrow }$ are called heat conductance rates and the heat conductivity (heat transport rate) is defined as $\Gamma \equiv k_{\downarrow }-k_{\uparrow }$ \cite{Kosloff2017TheQH}. Note that, heat transport is not the only source of irreversibility in a quantum heat engine. Non-commutativity of the Hamiltonians at different times results in an additional dissipation channel, dubbed "quantum friction"~\cite{PhysRevE.65.055102}.\\

Thermalization times are often neglected in the analyses of quantum heat engines, since they are assumed to be much shorter than the compression and expansion times.~\cite{PhysRevLett.112.030602,PhysRevE.98.032121,PhysRevE.99.022110,PhysRevE.99.032108,e18050168}. However, this assumption requires heat transport rates to be large, a circumstance for which it is shown that the optimal power production happens when the cycle times are vanishingly small~\cite{10.1119/1.18197,Rezek_2006,Kosloff2017TheQH}.\\

Another line of reasoning to justify small thermalization times in the isochoric process is that, for the frequency to stay constant (hence, the Hamiltonian to be static), the duration of the system-bath interaction must be shorter than the duration of the isentropic process. Meanwhile, assuming that there exists a considerable energy exchange between the system and the bath or that the system equilibrates to the temperature of the bath, controlling the
interaction time is equivalent to adjusting the system-bath coupling. However, a large system-bath coupling also implies a large heat transport rate. Additionally, in a thermalization map, which follows a GKLS type master equation, there is an implicit assumption that the thermalization time (which scales proportional to the inverse of the system-bath interaction energy), is longer than the relaxation dynamics of the bath and the unperturbed dynamics of the
system~\cite{1979}. Therefore, for a realistic and consistent thermodynamic analysis of a quantum heat engine in its full generality, it is important to take into account the effect of thermalization times without imposing additional assumptions.\\

In order to see the effect of the duration of isochoric strokes, we take the total cycle time to be $\tau_{tot} = \tau_{12} + \tau_{23} + \tau_{34} + \tau_{41}$. We assume that $\tau_{adi}=\tau_{12} = \tau_{34}$ and $\tau_{iso}=\tau_{23} = \tau_{41}$. The engine power can be expressed as,
\begin{equation}\label{engine.power}
  P = -\frac{\left\langle W_{12}\right\rangle + \left\langle W_{34}\right\rangle}{\tau_{\text{tot}}}
\end{equation}
in which $\tau_{tot}=2\tau_{adi}+2\tau_{iso}$. For an ideal engine, efficiency becomes
\begin{equation}\label{engine.efficiency}
  \eta = -\frac{\left\langle W_{12}\right\rangle + \left\langle W_{34}\right\rangle}{\left\langle Q_{23}\right\rangle}.
\end{equation}
Due to the fact that the ideal adiabatic and isochoric strokes are quasi-static, upon implementing a full cycle in finite time, the state
the working medium of the engine doesn't precisely return to the initial state from which the cycle started. Therefore, for a complete study of a finite time quantum heat engine a limit cycle analysis has to be taken into account and the thermodynamic behavior of the engine in this mode of operation needs to be
addressed~\cite{Kosloff2017TheQH,PhysRevE.97.062153,PhysRevE.70.046110}.\\

Existence of a limit cycle for a quantum heat engine can be argued in very general terms. A general quantum process is described by a completely positive trace preserving (CPTP) map. The quantum channel for the entire cycle is a composition of four CPTP maps,
\begin{equation}\label{cyclemap}
  \varepsilon(\hat{\rho})= \varepsilon_{4\rightarrow1}\circ\varepsilon_{3\rightarrow4}\circ\varepsilon_{2\rightarrow3}\circ\varepsilon_{1\rightarrow2}(\hat{\rho})
\end{equation}
which makes itself CPTP. Due to the Brouwer fixed-point theorem, every such channel should have at least one density operator fixed point $\rho^*$ such that~\cite{watrous_2018}
\begin{equation}\label{limitcyc}
  \varepsilon(\hat{\rho}^*) = \hat{\rho}^*.
\end{equation}
Lindblad showed that relative entropy (KL-divergence) between a state and a reference state is contractive under CPTP
maps~\cite{Lindblad1975}. Later, Petz proved that a variety of metrics (including the Bures metric) are monotone (contractive) under CPTP maps~\cite{PETZ199681}. These theorems can be expressed mathematically as 
\begin{align}\label{monotone}
  D_{\text{KL}}(\varepsilon(\hat{\rho})||\varepsilon(\hat{\rho}_{\text{ref}})) \leq D_{\text{KL}}(\hat{\rho}||\hat{\rho}_{\text{ref}}),\\
  D(\varepsilon(\hat{\rho}),\varepsilon(\hat{\rho}_{\text{ref}})) \leq D(\hat{\rho},\hat{\rho}_{\text{ref}})
\end{align}
Here, $D_{\text{KL}}$ is the relative entropy (which is not a metric but a divergence function) and $D$ is a metric on the statistical manifold of the density operators. According to these theorems, the two states become less and less distinguishable upon successive application of the map $\varepsilon$. If the fixed point of the CPTP map is unique, the density operator will monotonically approach the steady-state via the dynamical map~\cite{Frigerio1977,Frigerio1978}. In this case, if we take $\hat{\rho}_{\text{ref}}=\hat{\rho}^*$,~\cref{limitcyc} and~\cref{monotone} imply that upon successive application of the quantum channel, the system monotonically approaches the fixed point.

\section{STA by CD-driving}
\label{sec:sta}
In $1\rightarrow 2$ and $3\rightarrow 4$ the quantum system is isolated and work is performed by changing the frequency between $\omega_c$ and $\omega_h$, which are fixed as design parameters. We now consider a counter-diabatic drive on the system by means of a control Hamiltonian added to $\hat{H}_0$, i.e.,
\begin{equation}\label{cd.ham.sta}
\hat{H}_{CD}(t) = \hat{H}_0(t) + \hat{H}_1(t)
\end{equation}
such that the resulting dynamics given by
\begin{equation}\label{eq.mo.sta}
   \partial_t{\hat{\rho}} = -\frac{i}{\hbar}[\hat{H}_{CD},\hat{\rho}]
\end{equation}
is identical to the adiabatic evolution under $\hat{H}_0$ alone. 
For a closed system, the general form of $\hat{H}_1$ reads~\cite{Berry_2009}
\begin{equation}\label{cd.drive.sta}
 \hat{H}_1(t) = i\hbar \sum_n (\ket{\partial_t n}\bra{n} - \braket{n}{\partial_t n}\ket{n}\bra{n})
\end{equation}
where $\ket{n}$ is the $n$-th eigenstate of $\hat{H}_0(t)$. This STA protocol requires that $\langle \hat{H}_1(0)\rangle = \langle \hat{H}_1(\tau)\rangle = 0$. Hence, the frequency satisfies the following initial and final conditions.
\begin{align}\label{bc.sta}
  \omega_t(0)=\omega_i,\quad \dot{\omega}_t(0)=0, \quad \ddot{\omega}_t(0)=0,\nonumber  \\
  \omega_t(\tau)=\omega_f,\quad \dot{\omega}_t(\tau)=0, \quad \ddot{\omega}_t(\tau)=0.
\end{align}
An interpolating ansatz which satisfies the mentioned boundary conditions is~\cite{PhysRevX.4.021013},
\begin{equation}\label{freq.ansazt.sta}
   \omega_t = \omega_i + (\omega_f - \omega_i)[10s^3-15s^4+6s^5]
\end{equation}
in which $s=t/\tau_{adi}$. For a time-dependent QHO, $\hat{H}_1$ in~Eq.(\ref{cd.drive.sta}) can be expressed as~\cite{Muga_2010}
\begin{equation}\label{qho.h1.sta}
\hat{H}_1 = -\frac{\dot{\omega}_t}{4\omega_t}(\hat{x}\hat{p}+\hat{p}\hat{x}).
\end{equation}
The total Hamiltonian governing the time evolution of the STA controlled QHO becomes
\begin{equation}\label{qho.hcd.sta}
\hat{H}_{CD} = \frac{\hat{p}^2}{2m}+\frac{m\omega_t^2\hat{x}^2}{2} -\frac{\dot{\omega}_t}{4\omega_t} (\hat{x}\hat{p}+\hat{p}\hat{x}).
\end{equation}
Hence, for a time dependent QHO, the total Hamiltonian is quadratic in $\hat{x}$ and $\hat{p}$ and can be considered a generalized QHO~\cite{PhysRevE.96.012133},
\begin{equation}\label{eff.hamiltonian.cd}
\hat{H}_{CD} = \hbar \Omega_t(\hat{b}_t^{\dagger}\hat{b}_t + 1/2),
\end{equation}
with an effective frequency $\Omega_t$ and the ladder operator $\hat{b}_t$ given by
\begin{eqnarray}\label{eff.freq.lad}
  \Omega_t &=& \omega_t\sqrt{1-\dot{\omega}_t^2/4\omega_t^4}; \\
  \hat{b}_t &=& \sqrt{\frac{m\Omega_t}{2\hbar}}(\zeta_t \hat{x} + \frac{i\hat{p}}{m\Omega_t}).
\end{eqnarray}
The work done during this process is given by
\begin{equation}\label{work.adi}
W = \int_{0}^{\tau_{adi}}\text{Tr}(\dot{\hat{H}}_0(t) \hat{\rho}(t)) dt
\end{equation}
The expectation value of the CD driving is defined as~\cite{PhysRevE.98.032121},
\begin{multline}\label{exp.val.cd}
\left\langle \hat{H}_1(t) \right\rangle = \left\langle \hat{H}_{CD}(t) \right\rangle - \left\langle \hat{H}_0(t) \right\rangle=
\\ \frac{\omega_t}{\omega_i} \left\langle \hat{H}_0(0) \right\rangle \left(\frac{\omega_t}{\Omega_t}-1\right). 
\end{multline}

In order to obtain an accurate thermodynamic description of this controlled process, one should also keep track of the energy cost of CD-driving. The energetic cost of CD driving can be given as the time average of $\left\langle \hat{H}_1(t) \right\rangle$~\cite{PhysRevE.98.032121}.
\begin{equation}\label{sta.cost}
C_{\text{STA}} = \frac{1}{\tau_{adi}}\int_{0}^{\tau_{\text{adi}}} \left\langle \hat{H}_1(t) \right\rangle dt.
\end{equation}

\section{STE by CD-driving}
\label{sec:stt}
In the previous section we briefly explained superadiabatic control protocol of an isolated quantum system which will be used to accelerate the unitary compression and expansion strokes of our quantum Otto engine. However, to be able to accelerate thermalization processes in quantum heat engines, one must implement a similar protocol for an open system. Such control methods are of utmost importance for designing quantum thermal devices within the context of finite time thermodynamics. In this section we will briefly explain a fast thermalization protocol for a driven open QHO which is presented in~\cite{Alipour2020shortcutsto,PhysRevResearch.2.033178}. For open quantum systems, eigenvalues of the density matrix are time dependent. The time evolution of the density matrix can be written in terms of changes in its eigenprojectors and changes in the eigenvalues. Such a trajectory map can be written as
\begin{equation}\label{eq.mo.ste}
   \dot{\hat{\rho}} = -\frac{i}{\hbar}[\hat{H}_{CD},\hat{\rho}] + \sum_{n}\partial_t \lambda_n(t)\ket{n_t}\bra{n_t}.
\end{equation}
It is known that (after a modification to make it norm preserving) we can cast \cref{eq.mo.ste} into a Lindblad-like form~\cite{Alipour2020shortcutsto},
\begin{equation}\label{lind.like.ste}
   \partial_t \tilde{\rho} = \sum_{mn}\gamma_{mn}(\tilde{L}_{mn} \tilde{\rho} \tilde{L}_{mn}^{\dagger} - \frac{1}{2}\{\tilde{L}_{mn}^{\dagger} \tilde{L}_{mn}, \tilde{\rho}\}),
\end{equation}
in which $\tilde{L}_{mn}$ and $\gamma_{mn}$ are the jump operators and the dissipation rates in the Lindblad-like master equation for a
unitarily equivalent trajectory $\tilde{\rho}$. In~\cite{Alipour2020shortcutsto} it is shown that, using the CD-driven STE framework, one can achieve the fast thermalization of a time-dependent QHO from an initial to a final thermal state. For an open and time-dependent QHO, we can write the trajectory map for the unitary transformed density matrix $\tilde{\rho}=\hat{U}_x \hat{\rho} \hat{U}_x^{\dagger}$ with $\hat{U}_x = \text{exp}(im\alpha_t \hat{x}^2/2\hbar)$ as~\cite{Alipour2020shortcutsto},
\begin{equation}\label{qho.ste.master}
\partial_t \tilde{\rho} = \frac{-i}{\hbar} \left[\frac{\hat{p}^2}{2m} + \frac{1}{2} m\tilde{\omega}_{CD}^2 \hat{x}^2,  \tilde{\rho}\right] - \gamma_t \left[\hat{x}, \left[\hat{x}, \tilde{\rho}\right]\right],
\end{equation}
in which $\tilde{\omega}_{CD}$ is the effective frequency and $\gamma_t$ is the effective dissipation rate. Such a procedure for shortcut to thermalization can be implemented by modulation of the driving frequency and the dephasing strength~\cite{Alipour2020shortcutsto,PhysRevResearch.2.033178}. For~\cref{qho.ste.master} we have~\cite{Alipour2020shortcutsto},
\begin{eqnarray}\label{qho.ste.param}
  \omega_{CD}^2 &=& \omega_t^2 - \alpha_t^2 - \dot{\alpha}_t; \\
  \gamma_t &=& \frac{m\omega_t}{\hbar}\frac{\dot{u}_t}{(1-u_t)^2}; \\
   u_t &=& e^{-\beta_t \hbar \omega_t}; \\
   \alpha_t &=& \zeta_t - \dot{\omega}/2\omega_t; \\
   \zeta_t &=& -\frac{\dot{\omega}}{2\omega_t} + \frac{\dot{u}_t}{1-u_t^2}.
\end{eqnarray}
Here, $\beta_t$ is the inverse temperature at time $t$. Similar to the STA case, shortcut to equilibrium is achieved for a general setup by requesting $\omega_t$ to follow \cref{freq.ansazt.sta} (with $s=t/\tau_{iso}$) and $\beta_t$ to conform to the ansatz~\cite{Alipour2020shortcutsto},
\begin{equation}\label{beta.ansazt.ste}
   \beta_t = \beta_i + (\beta_f - \beta_i)[10s^3-15s^4+6s^5],
\end{equation}
in which $s=t/\tau_{iso}$. For this protocol, an entropy based definition for the infinitesimal thermal energy and work can be calculated as given in~\cite{PhysRevA.105.L040201}.
\begin{align}
    \dbar\mathcal{Q} &= \dbar Q - \dbar W_{CD} = \text{Tr}\left[\tilde{\mathcal{D}}_{CD}(\tilde{\rho}) \hat{H}_0 \right]dt;\label{heat.ste}\\
    \dbar\mathcal{W} &= \dbar W + \dbar W_{CD} = \text{Tr}\left[\tilde{\rho}(\dot{\hat{H}}_0-i\left[\hat{H}_0,\tilde{H}_{CD} \right]) \right]dt\label{work.ste}
\end{align}
in which $\tilde{H}_{CD}$ is the transformed CD Hamiltonian governing the coherent evolution and $\tilde{\mathcal{D}}_{CD}=-\gamma_t \left[\hat{x},\left[\hat{x},\tilde{\rho}\right]\right]$ is the transformed dissipator contributing to the coherent and incoherent evolution in~\cref{qho.ste.master} respectively. $\dbar W_{CD}=-i\text{Tr}[[\hat{H},\tilde{H}_{CD}] \tilde{\rho}]dt$ is the dissipative work, due to the CD evolution along the state trajectory. $\dbar\mathcal{Q}$ and $\dbar\mathcal{W}$ are heat and work calculated in the instantaneous eigenbasis of $\tilde{\rho}$. In~\cref{heat.ste} and~\cref{work.ste}, $\dbar Q$ and $\dbar W$ signify the infinitesimal changes in the conventional definitions of heat $Q=\text{Tr}[\dot{\hat{\rho}}\hat{H}_0]$ and work $W=\text{Tr}[\hat{\rho}\dot{\hat{H}}_0]$, resulting from Spohn separation~\cite{10.1063/1.523789}, in which the coherent (dissipative) changes of the internal energy due to the master equation is phenomenologically labelled as work (heat).\\

However, it is argued in~\cite{PhysRevA.105.L040201} that for a general evolution of an open quantum system, the contribution of changes in the internal energy which one can label as heat must also coincide with the contribution associated with changes in the entropy. In general, Spohn seperation doesn't necessarily respect this condition. As a resolution to this issue, the contribution to the changes in the internal energy associated with changes in the eigenvalues of the density matrix is called heat and the contribution associated with the changes in the eigenprojectors is called work. For the CD-driven harmonic oscillator driven to thermalization, these contributions take the form given in~\cref{heat.ste} and~\cref{work.ste} as discussed in~\cite{PhysRevA.105.L040201}. Therefore, for the driven system described in this section, the actual values for heat ($\mathcal{Q}$) and work ($\mathcal{W}$) during the process can be found by integrating~\cref{heat.ste} and~\cref{work.ste} throughout the trajectory. $\dbar W_{CD}$ quantifies the environment-induced dissipative work (positive or negative) due to CD driving of the open quantum system evolving for a time $dt$. More specifically, it is argued that for a system initialized in an equilibrium state and whose dynamics has a unique equilibrium steady state, the amount of $\dbar W_{CD}$ quantifies the energetic cost associated with the difference between the real trajectory and the quasistatic one~\cite{PhysRevA.105.L040201}. Therefore, the magnitude of the dissipative work at each infinitesimal time step can be associated with the control cost for its corresponding infinitesimal evolution along the trajectory. Since during an isochoric process the Hamiltonian is static ($\dot{\hat{H}}_0=0$) we can define the integral of the absolute value of $\dbar W_{CD}$ through the state trajectory during this process as the total cost function
\begin{equation}\label{ste.cost}
C_{\text{STE}} = \int_{0}^{\tau_{\text{iso}}} \left|\text{Tr}[[\hat{H}_0,\tilde{H}_{CD}] \tilde{\rho}]\right|dt.
\end{equation}
This cost is not defined operationally but based on the arguments that are given, any driving protocol must expend at least the amount of energy given in~\cref{ste.cost} to drive the system for the duration of the isochore. Therefore it can be thought of as a lower bound on any operational cost that one can define for such a process.

\section{Otto engine with CD driven STA and STE}
\label{sec:otto_sta_ste}
Using the frameworks explained in the previous sections, we propose a CD driven Otto engine with a QHO as a working medium. In \cref{otto.schm} we boost the engine using CD driven STA protocols for expansion and compression and using a CD driven STE protocol for the hot isochore. The cold isochore is modeled by a Markovian master equation and no control Hamiltonian is applied during this step.\\
%%%%%%%%%%%%%%%%%%%%%%%% figure is placed here just to be displayed properly in the article
\begin{figure*}[t!]
\centering
\includegraphics[width=\linewidth]{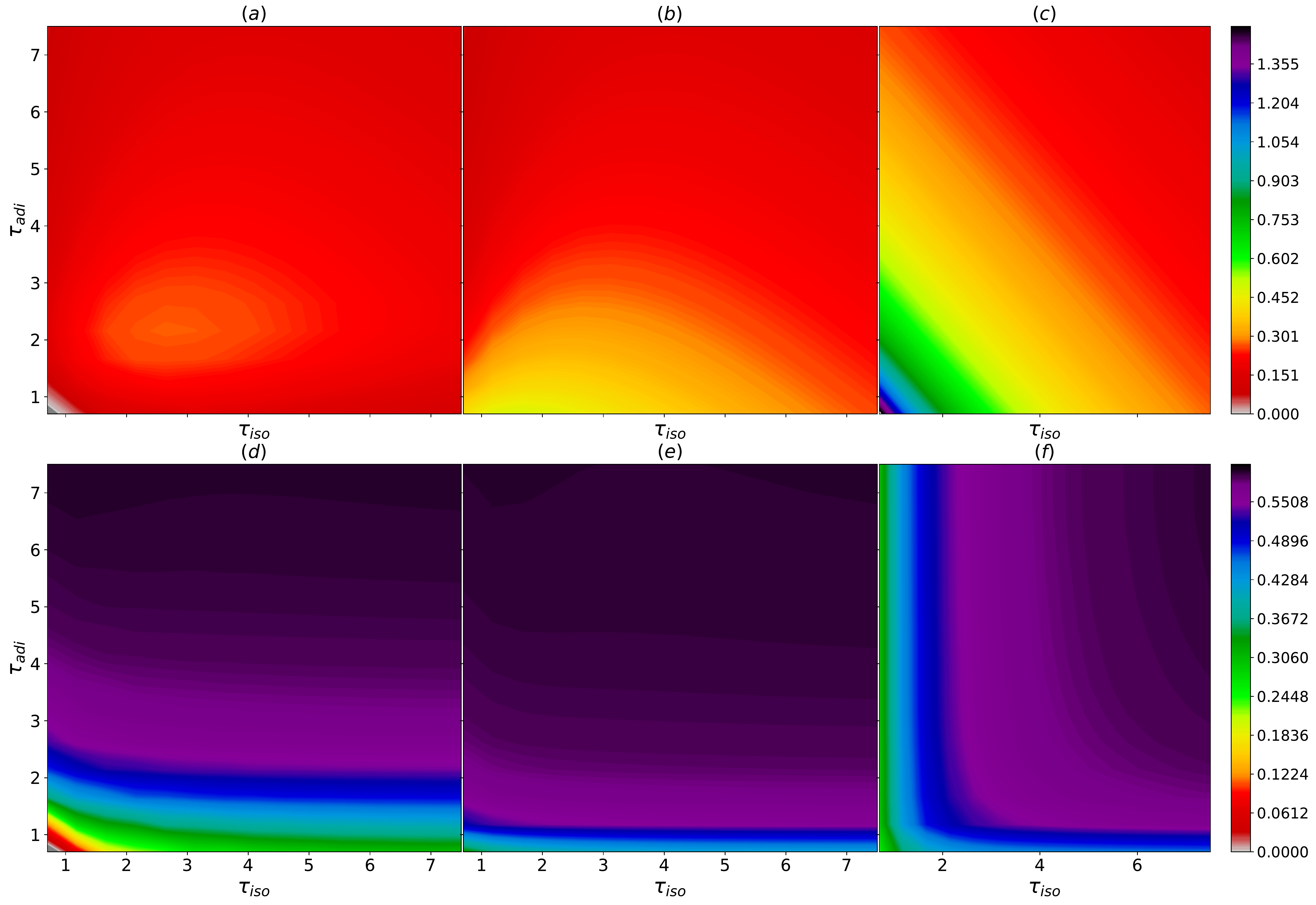}
\caption{Power and efficiency of the Otto engine as a function of the duration of the isochoric and adiabatic branches. Top row: Power vs $\tau_{adi}$ and $\tau_{iso}$ for the (a) UNA engine; (b) STA engine; (c) ST{\AE} engine. Bottom row: Efficiency vs $\tau_{adi}$ and $\tau_{iso}$ for the (d) UNA engine; (e) STA engine; (f) ST{\AE} engine. Negative power and efficiency results for the finite time Otto engine are greyed out.\label{pow_eff_all}}
\end{figure*}
%%%%%%%%%%%%%%%%%%%%%%%%%%%%%%%%%%%%%%%%%%%%%%%%%%

Adiabatic evolution implies $\beta_f \omega_f = \beta_i \omega_i$. Hence, after an adiabatic driving for the compression (expansion) stroke, the evolution of the system leads to a density matrix whose statistical distance is close to the thermal density matrix at $\omega_2 = \omega_h$ and $T_2 = (\omega_h/\omega_c)\times T_c$ ($\omega_4 = \omega_c$ and $T_4 = (\omega_c/\omega_h)\times T_c$). We can quantify this statistical distance using fidelity. Also, due to the fact that for an isochoric process the frequency must be kept constant, we will use $\omega_t=\omega_h$ throughout the CD-driven hot isochoric stroke, which means that in order to achieve STE we will only consider modulation of the dephasing strength.\\

We distinguish between the thermodynamic efficiency ($\eta^{th}$) of the control engine, which doesn't take into account the control costs and is given by an expression similar to~\cref{engine.efficiency}, and the operational efficiency ($\eta^{op}$) which considers the energetic costs. The operational efficiency for the CD-driven Otto engine can be calculated taking into account the costs for STA and STE protocols.
\begin{equation}\label{op.efficiecy}
  \eta^{op} = -\frac{W_{12} + W_{34}}{\mathcal{Q}_{23}+C_{adi}^{1\rightarrow 2} + C_{iso}^{2\rightarrow 3} + C_{adi}^{3\rightarrow 4}}.
\end{equation}

In~\cref{op.efficiecy}, $W_{12}$ and $W_{34}$ are calculated using~\cref{work.adi} and $\mathcal{Q}_{23}$ is calculated by integrating~\cref{heat.ste}. $C_{adi}^{1\rightarrow 2}$, $C_{iso}^{2\rightarrow 3}$ and $C_{adi}^{3\rightarrow 4}$ are the driving costs for the compression, cold isochore and expansion strokes and are defined in the previous sections. Power output is calculated using a similar expression given in~\cref{engine.power} by substituting $W_{12}$ and $W_{34}$ into the equation.

\section{Results}
\label{sec:result}
In this section, we compare UNA, STA, and ST{{\AE}} Otto engines in terms of power and efficiency. To this end, we choose $\omega_c = \omega_1 = \omega_4 = 1$ and $\omega_h = \omega_2 = \omega_3 = 2.5$. We assume the durations of the two adiabatic strokes to be identical ($\tau_{adi}$), and similarly for the two isochores ($\tau_{iso}$). We vary $\tau_{adi}$ and $\tau_{iso}$ independently between 0.7 and 7.5. The temperatures for the cold and hot bath are taken to be $T_c=T_1=1$ and $T_h=T_3=10$, respectively. We further assume that the heat conductivities for the hot and cold isochores are equal, i.e.,  $\Gamma_h=\Gamma_c=0.22$. All numerical values are given in appropriate units (energy in units of $\hbar\omega_c$, temperature in units of $\hbar\omega_c/k_B$, etc.). The parameter ranges were chosen such that all relevant operational regimes were accessible in the analysis. Throughout this section, all the calculated efficiencies are operational except when it is directly stated otherwise. For convenience we drop the "op" superscript for the operational efficiency. We have used the open source package QuTiP for our calculations~\cite{qutip}.\\

\cref{pow_eff_all} compares both the power output of the three engines (top panel) and their efficiencies (bottom panel) as a function of the stroke durations, for a single cycle of operation starting from the thermal state ($\omega_c,T_c$) which is indicated as "1" in~\cref{otto.schm}. It is evident from~\cref{pow_eff_all}(a) that for small values of $\tau_{adi}$ and $\tau_{iso}$ the UNA engine results in a negative power output and hence, does not behave as a proper heat engine. However, in STA~\cref{pow_eff_all}(b) and ST{\AE}~\cref{pow_eff_all}(c) engines, a positive power output is produced even for small stroke times. It should be noted that we have not considered stroke times smaller than 0.7 because for the parameter regime that we chose, this results in trap inversion (with an infinite STA driving cost). The maximum power output of the ST{\AE} and STA engines occur for the smallest stroke times however for the uncontrolled engine maximum power occurs for an intermediate time $\tau_{adi}=\tau_{iso}\approx 2.3$, after which the power output declines. The figures demonstrate that the maximum power is larger for ST{\AE} engine than the STA-only engine, which itself yields higher power than the finite-time engine with no CD-driving. 
When cycle time is dominated by either isochoric or adiabatic processes the power output of the uncontrolled engine is low and, as expected, the STA engine yields higher a power output when the cycle time is dominated by the isochoric processes. However, ST{\AE} engine outputs a relatively large values for power whether the contribution of the isochoric processes is larger for the cycle time than the adiabatic ones or vice-versa. For large values of the cycle time, the difference between the power outputs of the three engines become negligible as expected.
\cref{pow_eff_all}(d) shows that in the parameter regime that we have considered, the UNA engine yields a negative value for efficiency as the cycle time becomes very small. Hence, it doesn't act as a proper heat engine. Comparing~\cref{pow_eff_all}(d),~\cref{pow_eff_all}(e) and~\cref{pow_eff_all}(f) we see that for small values of $\tau_{adi}$, the efficiency is comparatively small for all three engines. However, STA and ST{\AE} engines give larger values for the efficiency, the former displaying a superior performance. For cycle times dominated by $\tau_{adi}$ and short durations of the isochoric branches, the ST{\AE} engine performs poorly as compared with the UNA and STA engines. As $\tau_{adi}$ and $\tau_{iso}$ become larger, the difference between the values of the efficiency of the three engines diminish.
%%%%%%%%%%%%%%%%%%%%%%%% figure is placed here just to be displayed properly in the article
\begin{figure}[htbp!]
\centering
\includegraphics[width=\linewidth]{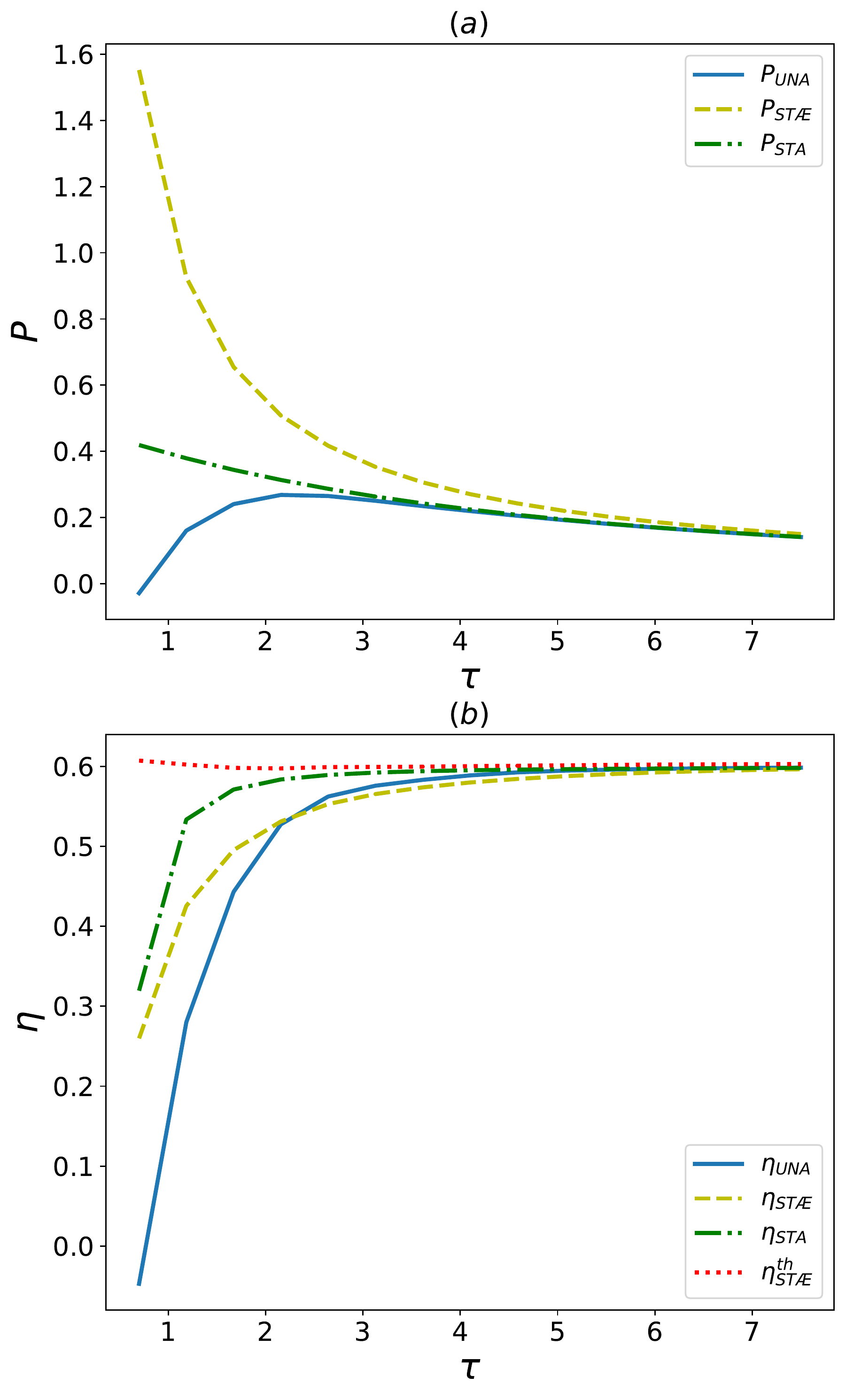}
\caption{Power and efficiency characteristics of the controlled and uncontrolled finite time Otto engines for an equally allocated time for their isochoric and adiabatic branches, $\tau_{iso}=\tau_{adi}=\tau$. (a) Power vs $\tau$: UNA engine (solid blue), STA engine (green dash-dotted), ST{\AE} engine (yellow dashed). (b) Efficiency vs $\tau$: UNA engine (solid blue), STA engine (green dash-dotted), operational efficiency (yellow dashed) and thermodynamic efficiency (red dotted) of the ST{\AE} engine. \label{power_efficiency_eqst}}
\end{figure}
%%%%%%%%%%%%%%%%%%%%%%%%%%%%%%%%%%%%%
To see finer details of the performance metrics of the three engines for a subset of the cycle times considered in~\cref{pow_eff_all}, in~\cref{power_efficiency_eqst} efficiency and power of the engines are presented for $\tau_{adi}=\tau_{iso}=\tau$. In~\cref{power_efficiency_eqst}(a) we see a dramatic difference between the power output of the ST{\AE} engine with the other two engines considered in this study specially for short cycle times. This figure shows a clear advantage of using CD-driving for thermalization and adiabaticity to enhance the power output. From~\cref{power_efficiency_eqst}(b) it is clear that for equal time allocated to isochoric and adiabatic branches, the largest to smallest efficiencies are obtained for the STA engine, the ST{\AE} engine and the UNA engine, respectively for smaller cycle times. For the intermediate cycle times efficiency of the UNA engine exceeds that of ST{\AE}. For larger cycle times the efficiency of the three engines approach the ideal value of $1-\omega_c/\omega_h=0.6$. Also, for the entirely of cycle times, the efficiency of all three engines is below the Carnot efficiency $1-T_c/T_h=0.9$ and Curzon-Ahlborn (CA) efficiency $1-\sqrt{T_c/T_h}\approx0.68$. Note that without considering the costs, the thermodynamic efficiency of the ST{\AE} engine equals the ideal value.\\
%%%%%%%%%%%%%%%%%%%%%%%% figure is placed here just to be displayed properly in the article
\begin{figure}[htbp!]
\centering
\includegraphics[width=\linewidth]{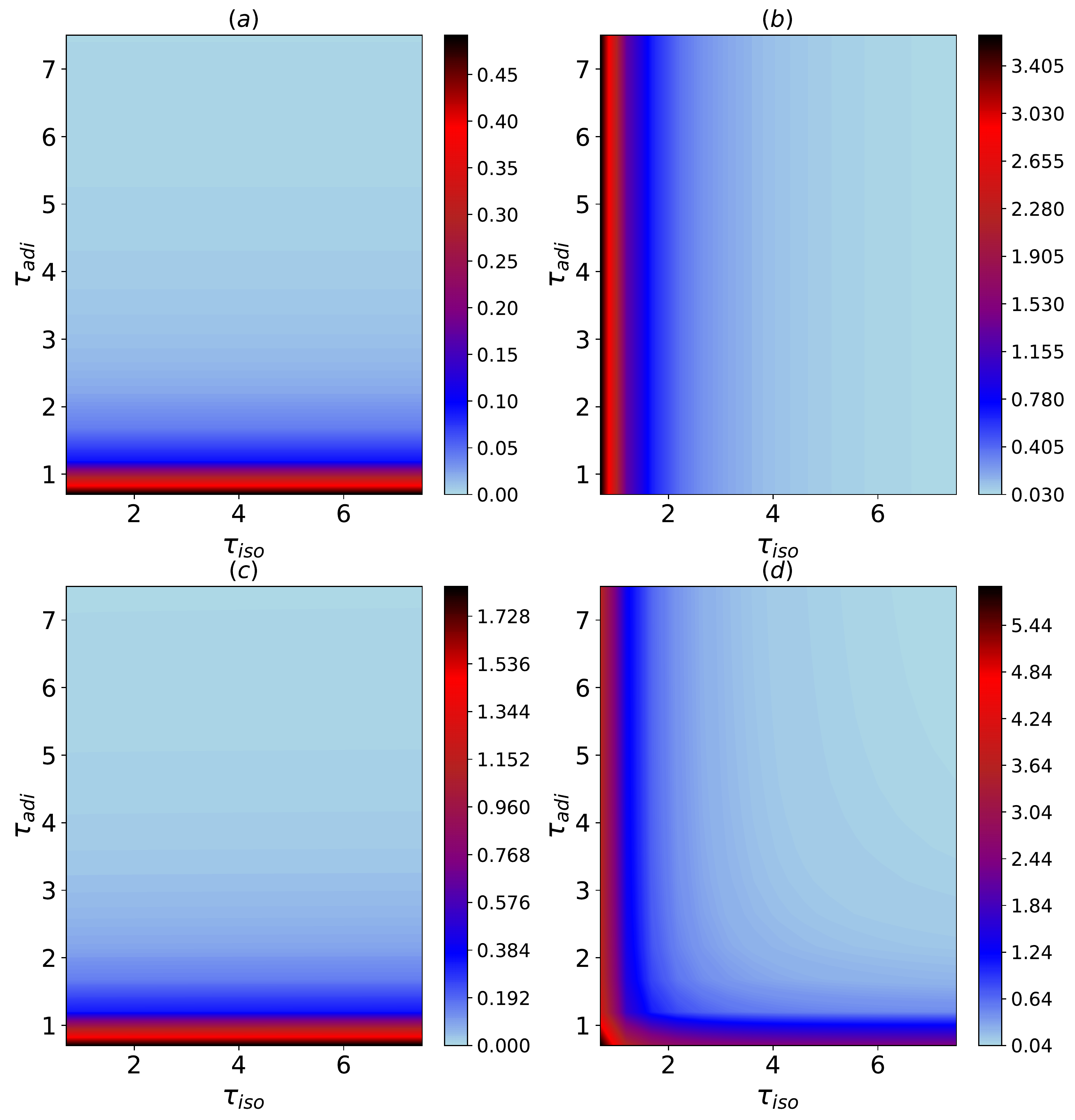}
\caption{Cost of the STA and STE driving for one cycle of the ST{\AE} engine vs $\tau_{adi}$ and $\tau_{iso}$. (a) Cost of the STA driving for adiabatic compression; (b) cost of the STE driving for hot isochore; (c) cost of the STA driving for adiabatic expansion; (d) total costs for one cycle. \label{costs_alltime}}
\end{figure}
In~\cref{costs_alltime} we can see the control costs for the ST{\AE} engine. As expected,~\cref{costs_alltime}(a) and~\cref{costs_alltime}(c) show that the energetic cost for CD-driving for adiabaticity decreases for longer $\tau_{adi}$, whereas in~\cref{costs_alltime}(b) we can see that the cost for driving the system to thermalization shrinks as we increase $\tau_{iso}$. Comparison between the costs for the controlled strokes in the ST{\AE} engine reveal that the maximum of the energetic control costs in increasing order are for the compression, expansion and thermalizaion strokes. Moreover, the expansion stage has more populations in excited levels as it starts after the heating stroke, in contrast to compression stage where the populations are less as it follows the cooling stroke. Accordingly, more energy needs to be spent by the external agent to control and stop the transitions during accelerated adiabatic transformation in the expansion stroke, a fact which was also stated in earlier studies~\cite{Li_2021}.
%%%%%%%%%%%%%%%%%%%%%%%% figure is placed here just to be displayed properly in the article
\begin{figure*}[t!]
\centering
\includegraphics[width=\linewidth]{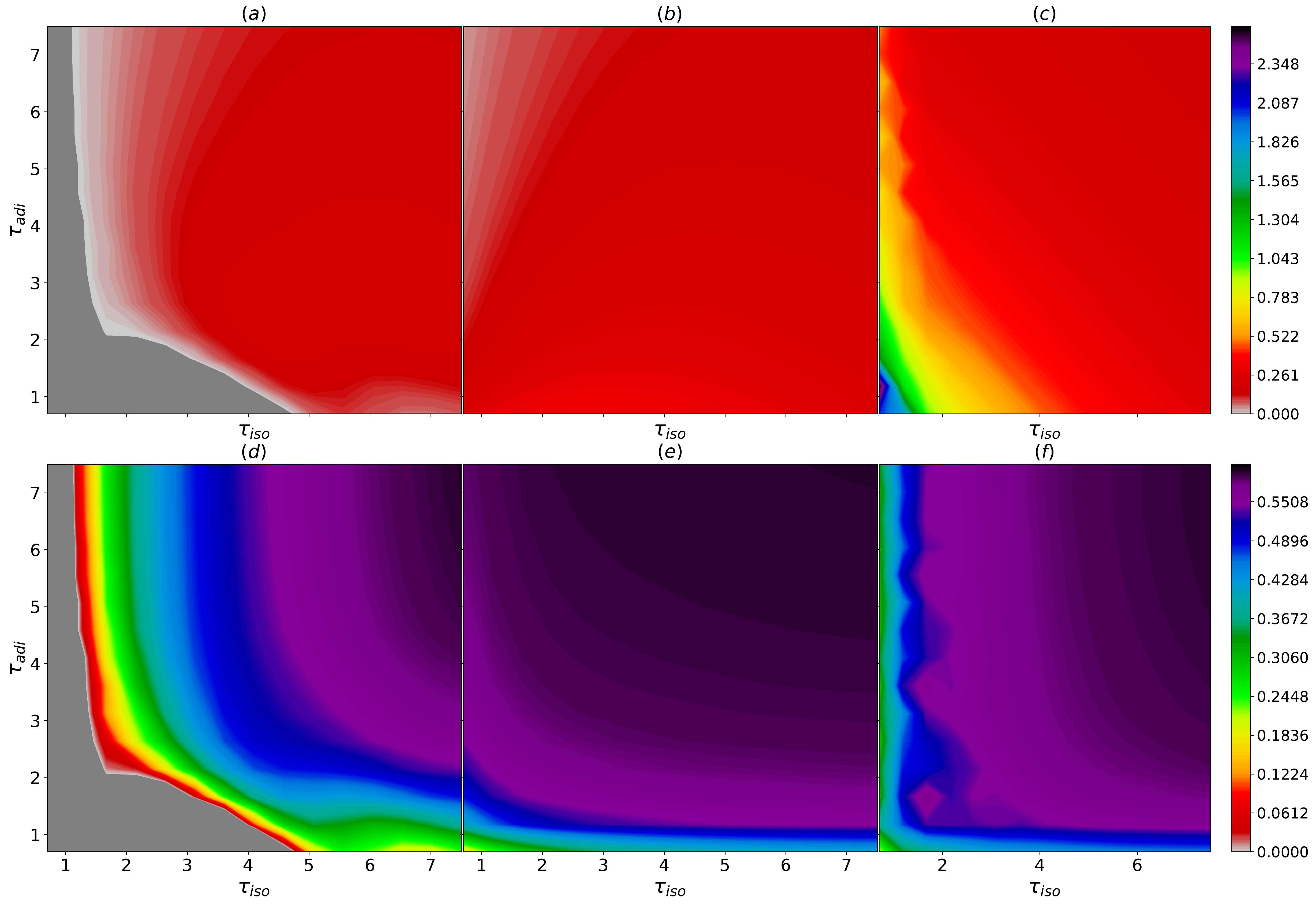}
\caption{Power and efficiency of the Otto engine at its limit cycle as a function of the duration of the isochoric and adiabatic branches. Top row: Limit cycle power vs $\tau_{adi}$ and $\tau_{iso}$ for the (a) UNA engine; (b) STA engine; (c) ST{\AE} engine. Bottom row: Limit cycle efficiency vs $\tau_{adi}$ and $\tau_{iso}$ for the (d) UNA engine; (e) STA engine; (f) ST{\AE} engine. Negative power and efficiency results for the finite time Otto engine are greyed out.\label{lc_pow_eff_all}}
\end{figure*}
%%%%%%%%%%%%%%%%%%
Next, we move on to the performance metrics of the engines in their respective limit cycles. Our numerical calculations show that all three engines reach a limit cycle throughout the cycle times that we have considered in this study. One has to note that in the limit cycle the state of the working medium at the end of each stroke do not necessarily coincide with the ones given in the ideal Otto engine as represented by~\cref{otto.schm}, unless the times allocated for each stroke are very large. Calculating the fidelity of the state of the working medium and the final state after each repetition of the cycle, we see observed that 7 cycles are enough to converge to the limit cycle of the engines for all the cycle times. In~\cref{lc_pow_eff_all} the results for the power and efficiency of the engines are shown.

From~\cref{lc_pow_eff_all}(a) we can observe that the UNA engine yields a negative power output for a considerably large subset of allocated times for the adiabatic and isochoric strokes. Comparing~\cref{lc_pow_eff_all}(a) and~\cref{pow_eff_all}(a), we see that the uncontrolled engine is much more unreliable in its limit cycle than when we only consider one cycle of operation. In comparison,~\cref{lc_pow_eff_all}(b) and~\cref{lc_pow_eff_all}(c) give a positive power output in their limit cycles for all $\tau_{adi}$ and $\tau_{iso}$. Moreover, in~\cref{lc_pow_eff_all}(a) we see that in its limit cycle, the UNA engine yields a vanishingly small power outputs for cycle times dominated by isochoric/adiabatic processes. On the other hand,~\cref{lc_pow_eff_all}(b) shows that the power output in the STA engine gets vanishingly small only for the cycle times dominated by adiabatic processes. However,~\cref{lc_pow_eff_all}(c) shows a clear advantage of using ST{\AE} engine, in which power output is much larger compared to the UNA and STA engines, specially when we consider isochoric or adiabatic process dominated cycle times. The figure shows that similar to the case of single cycle, for the STA{\AE} engine in its limit cycle, the power output is higher for smaller values of $\tau_{adi}$ and $\tau_{iso}$.
\begin{figure}[htbp!]
\centering
\includegraphics[width=\linewidth]{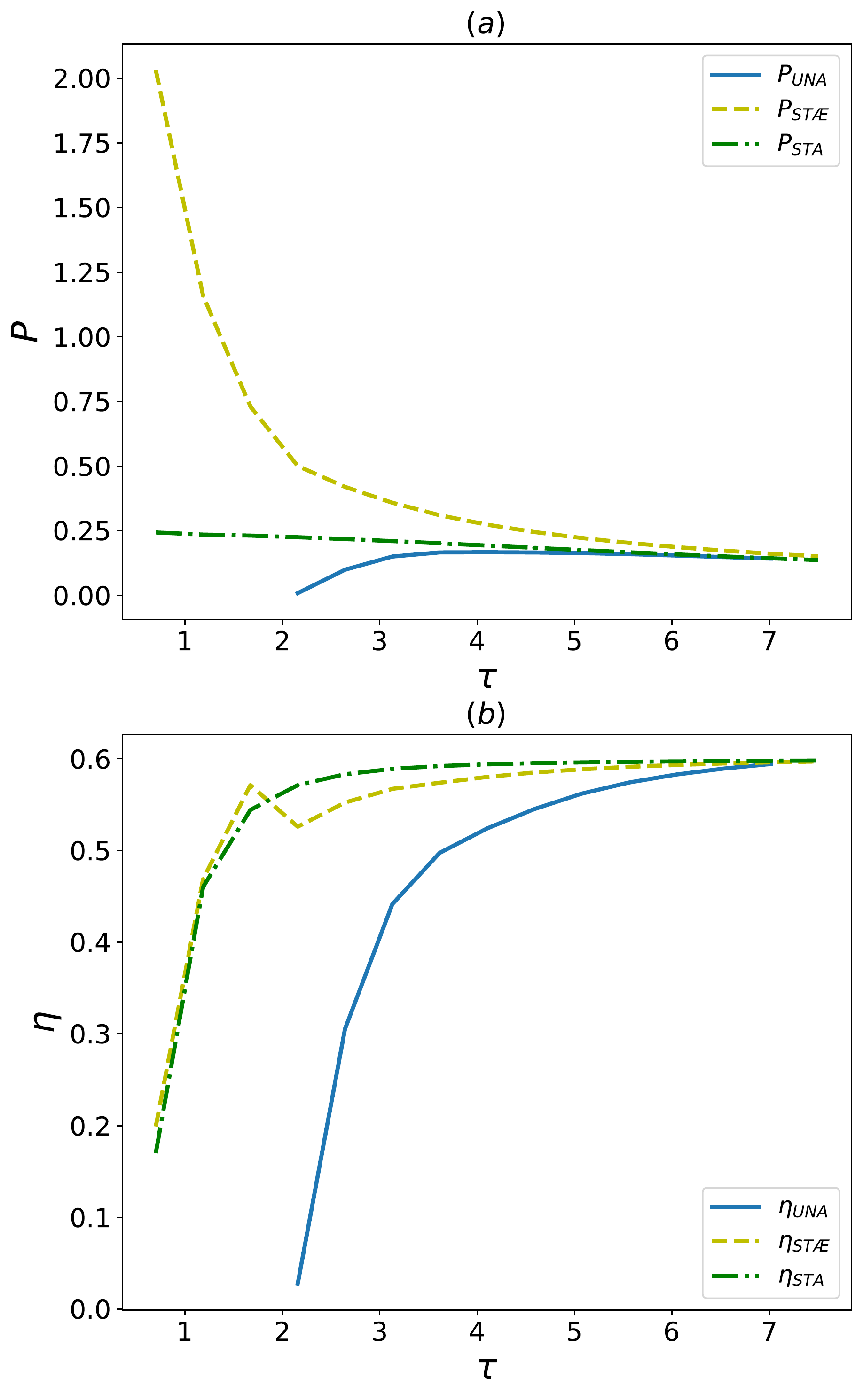}
\caption{Limit cycle power and efficiency characteristics of the controlled and uncontrolled finite time Otto engines for an equally allocated time for their isochoric and adiabatic branches, $\tau_{iso}=\tau_{adi}=\tau$. (a) Limit cycle power vs $\tau$: UNA engine (solid blue), STA engine (green dash-dotted), ST{\AE} engine (yellow dashed) of the engine with STA on its adiabatic strokes and STE on its hot isochore. (b) Limit cycle efficiency vs $\tau$: UNA engine (solid blue), STA engine (green dash-dotted), ST{\AE} engine (yellow dashed). Negative power and efficiency results for the finite time Otto engine are not displayed.\label{lc_power_efficiency_eqst}}
\end{figure}
The grey areas in~\cref{lc_pow_eff_all}(d) show that for a considerable section of the considered allocated times for the isochoric and adiabatic strokes, the uncontrolled engine doesn't act as a proper heat engine. In its limit cycle the efficiency of the UNA engine is low compared to the ST{\AE} and the STA engines except when $\tau_{iso}$ and $\tau_{adi}$ are both large. Comparing~\cref{lc_pow_eff_all}(e) and~\cref{lc_pow_eff_all}(f) we see that surprisingly, for cycles with small $\tau_{iso}$, the efficiency of the ST{\AE} engine is smaller than that of the STA engine except when $\tau_{adi}$ is also small. Although, this doesn't mean that the STA-only engine is superior to the ST{\AE} engine for the mentioned cycle time allocations because in this case, as seen in~\cref{lc_pow_eff_all}(b) the power output of the STA-only engine becomes much smaller than the ST{\AE} engine. An interesting aspect of the ST{{\AE}} protocol limit cycles is that the performance is non-monotonous in $\tau_{adi}$. As a result, high-power/high-efficiency islands appear, such as the one observed around $\tau_{adi}=\tau_{iso}\approx 1.6$ in \cref{lc_pow_eff_all}(f).
\begin{figure}[htbp!]
\centering
\includegraphics[width=\linewidth]{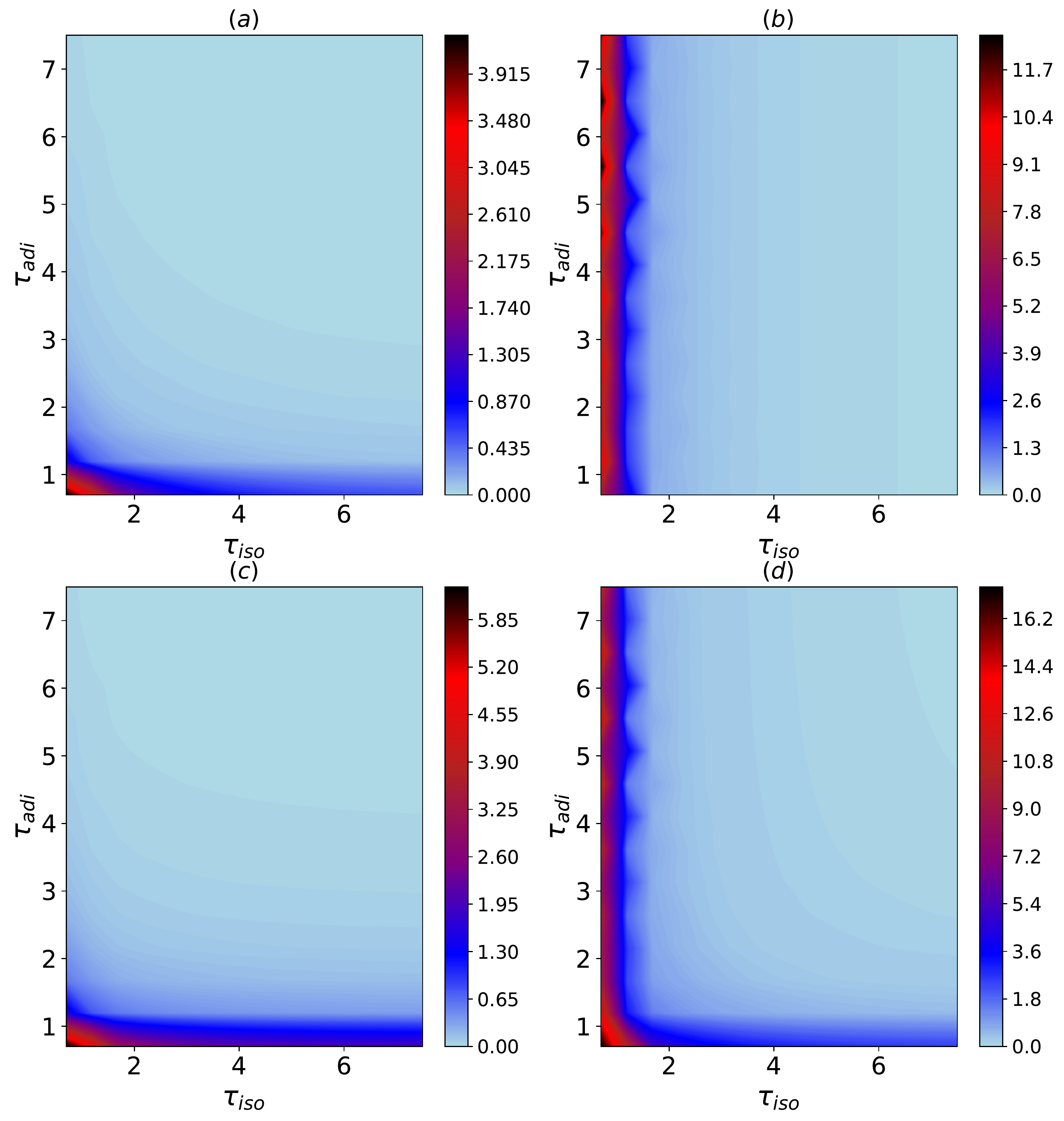}
\caption{Cost of the STA and STE driving for the limit cycle of the ST{\AE} engine vs $\tau_{adi}$ and $\tau_{iso}$. (a) Cost of the STA driving for adiabatic compression; (b) cost of the STE driving for hot isochore; (c) cost of the STA driving for adiabatic expansion; (d) total costs for one cycle. \label{lc_costs_alltime}}
\end{figure}
In order to make a more detailed comparison of the performance metrics of the engines for a subset of the stroke times considered in this study, in~\cref{lc_power_efficiency_eqst} we present efficiency and power of the heat engines for $\tau_{adi}=\tau_{iso}=\tau$. In~\cref{lc_power_efficiency_eqst}(a) we can see that the ST{\AE} engine yields a much higher power output than the STA and UNA engines. Comparing~\cref{power_efficiency_eqst}(a) and~\cref{lc_power_efficiency_eqst}(a) we see that in their limit cycle, the power of the STA and UNA engines get reduced, as compared with the case of single cycle, throughout the cycle times. However, for the ST{\AE} engine the power output for the limit cycle is higher than that of the single cycle for the entirety of the considered cycle durations. As shown in~\cref{lc_power_efficiency_eqst}(b), in its limit cycle, the efficiency of the ST{\AE} engine is no longer monotonic in the cycle time. The difference between the efficiencies of the uncontrolled engine and the controlled engines is more dramatic in their limit cycles and their efficiency values don't approach each other to the ideal value of $1-\omega_c/\omega_h=0.6$ unless the cycle time is very large. The results show that the ST{\AE} engine has higher efficiency than the STA engine, for smaller cycle times. However, after a certain value for the stroke durations, efficiency of the STA engine exceeds that of the ST{\AE} engine. Inspecting~\cref{power_efficiency_eqst}(a) and~\cref{power_efficiency_eqst}(b) we see that there exists a time interval for which the ST{\AE} engine achieves higher efficiency and power than the STA engine.\\

In~\cref{lc_costs_alltime} the costs for CD-driving at the limit cycle for the ST{\AE} engine are presented. Similar to the case of single cycle, we can see that the cost of the hot isochoric and adiabatic strokes are highest for short $\tau_{iso}$ and $\tau_{adi}$ respectively. Additionally, for short cycle times in the limit cycle the share of the thermalization cost is the highest in the total control cost similar to the case of single cycle. Again, as $\tau_{adi}$ and $\tau_{iso}$ is increased, the control costs of the adiabatic and hot isochoric strokes decreases. However, for small $\tau_{adi}$ in the limit cycle, the cost for CD-driving during the expansion and compression strokes becomes relatively small for larger $\tau_{iso}$ as opposed to what we see for a single cycle.\\

Comparing~\cref{costs_alltime} and~\cref{lc_costs_alltime} we see that for the limit cycle, the control costs tend to increase as compared to the single cycle case. The difference between the behavior of the cost functions for a single cycle and limit cycle can be explained by the fact that for short cycle durations, specifically for a short cold isochoric stroke, upon completion of a single cycle the system doesn't return to the thermal state at $T_c$ and $\omega_c$. Additionally, non-commutativity of the system Hamiltonian at different times and also non-commutativity of the system and control Hamiltonians will induce diabatic transition of the populations and also generate coherence in the system's density matrix in the energy eigenbasis. It is evident from the definition of the cost functions~\cref{sta.cost} and~\cref{ste.cost} that both the diagonals and off-diagonals of the density matrix influences the cost value along their respective processes. Therefore due to the mentioned diabatic transitions and generated coherences, which are more pronounced for smaller cycle durations, in the limit cycle the total control cost in the limit cycle will be larger than that of a single cycle.\\

\begin{figure*}[t!]
\centering
\includegraphics[width=\linewidth]{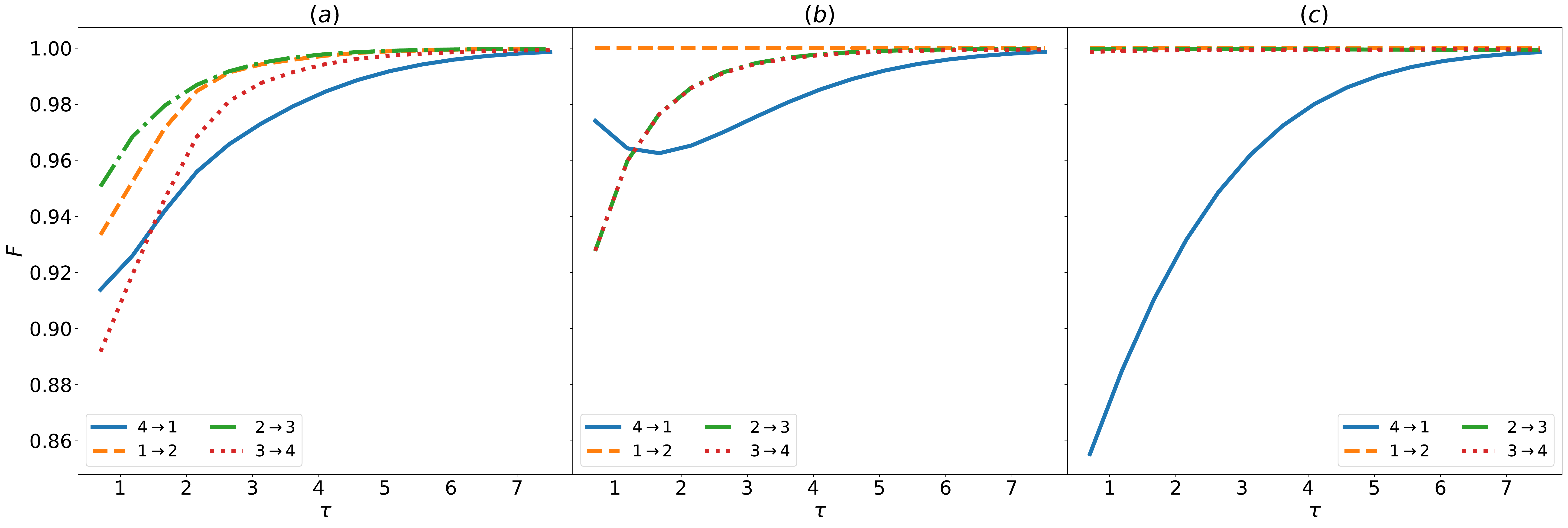}
\caption{Fidelity between the density matrix at the end of each stroke and their respective counterparts for an ideal quasistatic quantum Otto engine vs the stroke duration $\tau=\tau_{adi}=\tau_{iso}$. The calculations are carried out for a single cycle for (a) UNA engine (b) STA engine (c) ST{\AE} engine. The solid blue, dot-dashed green, dashed orange and dotted red curves are for the cold isochore, hot isochore, compression and expansion strokes respectively.\label{fid_onecyc}}
\end{figure*}

\begin{figure*}[t!]
\centering
\includegraphics[width=\linewidth]{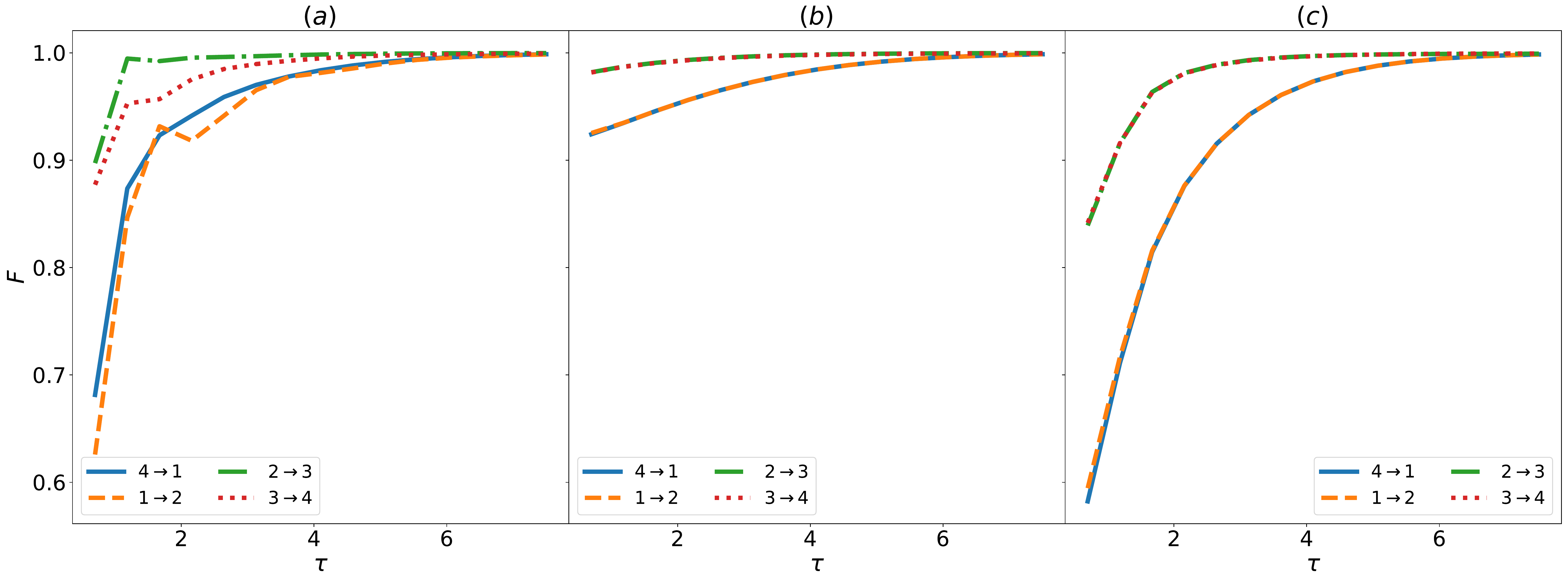}
\caption{Fidelity between the density matrix at the end of each stroke and their respective counterparts for an ideal quasistatic quantum Otto engine vs the stroke duration $\tau=\tau_{adi}=\tau_{iso}$. The calculations are carried out for the limit cycle for (a) UNA engine (b) STA engine (c) ST{\AE} engine. The solid blue, dot-dashed green, dashed orange and dotted red curves are for the cold isochore, hot isochore, compression and expansion strokes respectively.\label{fid_limcyc}}
\end{figure*}

\cref{fid_onecyc} shows the fidelity between the density matrix at the end of each stroke and their counterparts in an ideal quasistatic quantum Otto engine for different time allocation of the cycle durations. The density matrices at the end of each stroke for an ideal quasistatic heat engine are thermal states at $(T_i,\omega_i)$ in which $i=1,2,3,4$. The results are shown for a single cycle and an equal time duration for the isochoric and adiabatic strokes. If the fidelity between the density matrix at the end of a finite-time stroke in a controlled heat engine and its corresponding quasistatic counterpart is one, it is an indicator that the quantum evolution governing the dynamics of the stroke has successfully achieved its control target. From~\cref{fid_onecyc}(a) we see that in the UNA engine none of the four branches yield a high fidelity unless the stroke times are very large. From~\cref{fid_onecyc}(b) we see that implementation of STA drives the density matrix at the end of the compression stroke to its corresponding quasistatic counterpart with almost perfect fidelity regardless of the stroke duration. It is worth mentioning that the same doesn't happen for the expansion stroke because the thermalization stroke doesn't thermalize the system for smaller duration of this branch and the control law for the expansion presumes that the temperature at the beginning of the stroke is $T_3=T_h$. From~\cref{fid_onecyc}(c) we see that all of the controlled strokes in the ST{\AE} achieve perfect fidelity regardless of their duration.\\
%%%%%%%%%%%%%%%%%%%%%%%% 
%%%%%%%%%%%%%%%%%%
In~\cref{fid_limcyc} we show the fidelity between the density matrix at the end of each stroke and their counterparts in an ideal quasistatic quantum Otto engine for limit cycle and an equal time duration for the isochoric and adiabatic strokes. Since the three engines under investigation have different cycle propagators, they will reach different limit cycles at different times. This fact is evident upon comparing the fidelity curves for the cold isochore process of the UNA, STA and ST{\AE} engines in~\cref{fid_limcyc} for different times. Also from~\cref{fid_limcyc}(b) and~\cref{fid_limcyc}(c), we see that in the limit cycle, non of the controlled branches in the STA and ST{\AE} engines achieve unit fidelity unlike what was the case for a single cycle as displayed in~\cref{fid_onecyc}(b) and~\cref{fid_onecyc}(c). As explained earlier, this is due to the diabatic populations transitions and accumulated coherence in the energy eigenbasis (which increase the cost in limit cycle) and the differences between the fidelities for a single cycle and limit cycle shows confirms this fact.

%%%%%%%%%%%%%%%%%%%%%%%% 
\begin{figure}[htbp!]
\centering
\includegraphics[width=\linewidth]{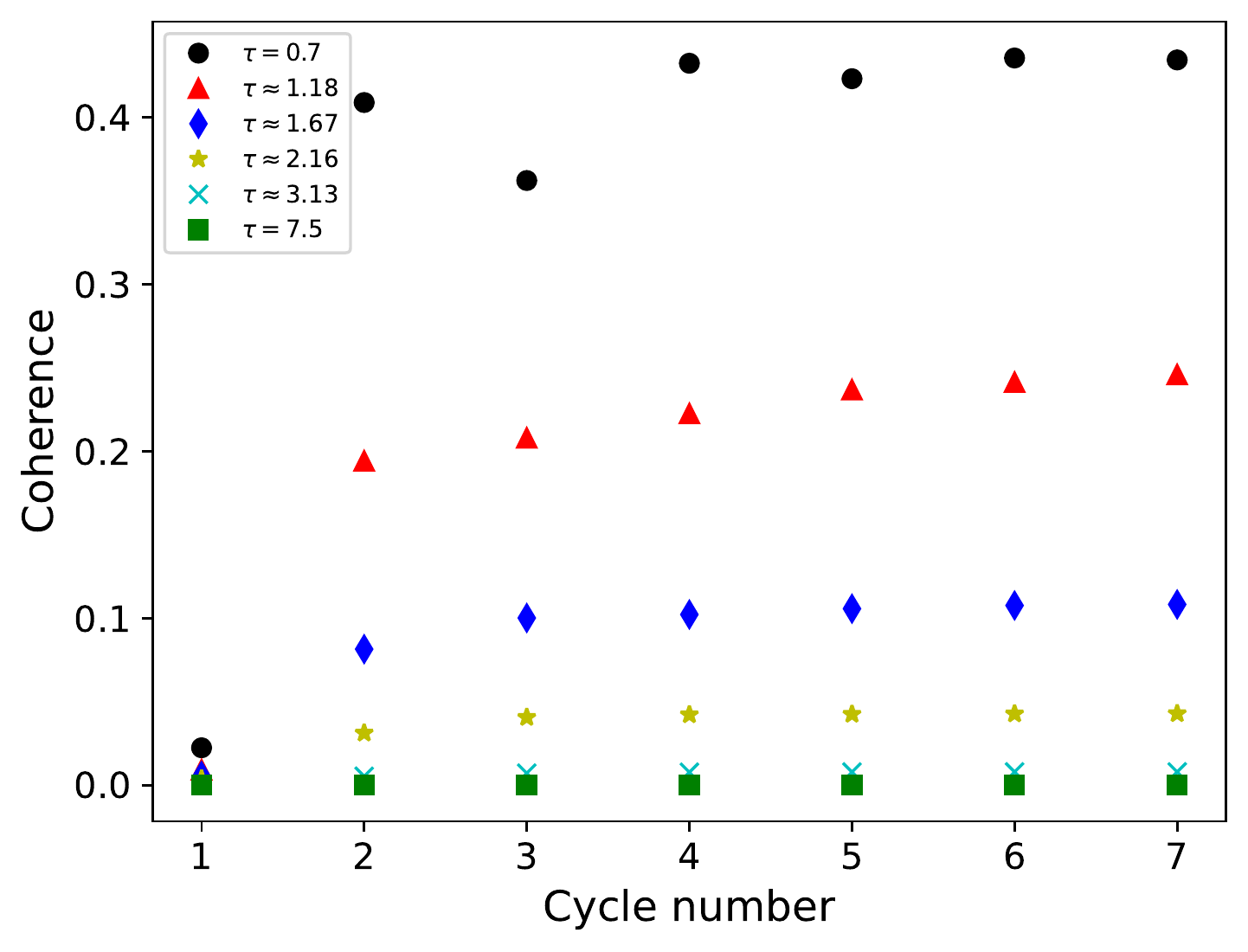}
\caption{l1-norm coherence of the density matrix at the end of cold isochore vs cycle number for different durations of cycle. $\tau=\tau_{adi}=\tau_{iso}$ for the ST{\AE} engine. \label{coherence_cyc}}
\end{figure}

%%%%%%%%%%%%%%%%%%
In~\cref{coherence_cyc} we provide the coherence accumulated in the density matrix for different $\tau=\tau_{iso}=\tau_{adi}$ calculated for each cycle number until the limit cycle is reached. To quantify coherence, l1-norm of coherence is used. The results show that as a general trend, for shorter cycle durations more coherence is accumulated in the density matrix. Also, as the cycle number increases, until the limit cycle is reached, the amount of coherence gets larger. The results displayed in~\cref{coherence_cyc} also demonstrate a clear reason based on the contribution of the off-diagonals of the density matrix that why the cost functions yield larger values at the limit cycle.

\section{Conclusion}
\label{sec:conclusion}
In conclusion, we studied the thermodynamic performance of a quantum harmonic Otto engine in finite time and compared three cases of Otto engine with no CD-driving, Otto engine with STA on the adiabatic strokes and Otto engine with STA on the adiabatic strokes and STE on the hot isochoric branch. We have included the costs of driving the engine for calculating the thermodynamic figures of merit. In our calculations we have considered two modes of operation. First we studied the thermodynamic performance criteria for a single cycle and then we calculated the same figures of merit for the engines running
in their respective limit cycles.\\

Our results for single cycle indicate that controlling the hot thermalization process in the Otto engine greatly increases the power output of the engine for the entirety of the considered stroke times, specifically for short cycle time durations. Taking the control costs into account, the ST{\AE} engine yields a lower efficiency than the STA and UNA engines for cycle times dominated by $\tau_{adi}$. At this mode of operation, the STA engine outperforms ST{\AE} engine in terms of efficiency. However, for short cycle durations the ST{\AE} engine results in a larger efficiency than the UNA engine. For a single cycle starting from the thermal state of the cold bath, the control costs for thermalization is larger than that of the compression and expansion strokes. In total, the control costs decline as we increase the time allocated for the isochoric and adiabatic branches. Our calculations show that for the UNA engine the parameter regime of the stroke time durations for which the power output and efficiency are negative greatly increases in its limit cycle as compared to only a single cycle of operation. For the CD-driven engines (both STA and ST{\AE}), the power output remains positive throughout the stroke times considered in this study.\\

For limit cycle the ST{\AE} engine displays a far more superior performance in terms of power output than the STA and UNA engines. The power for the ST{\AE} engine in this mode of operation increases unlike the STA and UNA engines compared to the single cycle case. Our numerical calculations show that in the limit cycle, the efficiency vs cycle time is not necessarily monotonic for cycle time for the ST{\AE}
engine. This is of practical importance because it means that one can design an ST{\AE} engine which has larger power output and efficiency compared to the STA engine. In its limit cycle, the thermalization cost dominates the control costs for the ST{\AE} engine, specially for small time allocation for the isochoric branches. However, similar to the single cycle case, the costs decline as the cycle time is increases. Due to the generation of coherence and also diabatic population transfer, the costs in the limit cycle are larger than that of a single cycle, specially for shorter stroke durations. In addition to illuminating the fundamental potential and energetic cost limitations of finite-time quantum heat engines with shortcuts to adiabaticity and equilibration, our results can be significant for their practical implementations with optimum power and efficiency.

%%%%%%%%%%%%%%%%%%%%%%%%%%%%%%%%%%%%%%%%
\acknowledgements
%%%%%%%%%%%%%%%%%%%%%%%%%%%%%%%%%%%%%%%

We gratefully acknowledge M.~T.~Naseem, B.~Çakmak, and O.~Pusuluk for fruitful discussions.

%%%%%%%%%%%%%%
% References %
%%%%%%%%%%%%%%
%\nocite{*}
%\bibliographystyle{apsrev4-2.bst}
%\bibliography{Refs_STA_STE_Otto}

\end{document}